\documentclass{amsart}

\newtheorem{theorem}{Theorem}[section]

\theoremstyle{definition}
\newtheorem{definition}[theorem]{Definition}

\theoremstyle{remark}
\newtheorem{remark}[theorem]{Remark}

\newcommand{\cA}{{\mathcal A}}
\newcommand{\cB}{{\mathcal B}}

\newcommand{\cH}{{\mathcal H}}

\newcommand{\cN}{{\mathcal N}}

\newcommand{\cS}{{\mathcal S}}

\newcommand{\Nn}{{\rm I\!N}} 
\newcommand{\Cn}{{\setbox0=\hbox{
$\displaystyle\rm C$}\hbox{\hbox
to0pt{\kern0.6\wd0\vrule height0.9\ht0\hss}\box0}}} 

\numberwithin{equation}{section}

\begin{document}
\begin{center}
\vspace*{15mm}
{\LARGE Quantum recurrences versus stability\footnote{
The support of Poland-South Africa Cooperation Joint project and (WAM) the support of
the grant No PBZ-MIN-008/PO3/2003 is gratefully acknowledged.}}\\
\vspace{2cm}
\textsc{{\large Louis E. Labuschagne} \\
Department of Maths, Applied Maths and Astronomy\\
University of South Africa\\
P.O.Box 392\\
0003 Pretoria, South Africa}\\
\textit{E-mail address:} \texttt{labusle@unisa.ac.za}\\
\vspace{1cm}
\textsc{{\large W{\l}adys{\l}aw A. Majewski}\\
Institute of Theoretical Physics and Astrophysics\\
Gda{\'n}sk University\\
Wita Stwosza~57\\
80-952 Gda{\'n}sk, Poland}\\
\textit{E-mail address:} \texttt{fizwam@univ.gda.pl}\\
\vspace{1cm}
\end{center}

\vspace*{3cm} \noindent \textsc{Abstract.} Consequences of quantum
recurrences on the stability of a broad class of dynamical systems
is presented.

\vspace{1cm}

\newpage


\section{Definitions, notations and stating the problem}
For any $C^*$-algebra $\cA$ let $\cA^+$ denote the set of all positive
elements in $\cA$. A~{\it state} on a~unital $C^*$-algebra $\cA$ is a
linear functional $\omega:\cA\to \mathbb{C}$ such that
$\omega(a)\geq 0$ for every $a\in \cA^+$ and $\omega(\mathbb{I})=1$
where $\mathbb{I}$ is the unit of $\cA$. By $\cS(\cA)$ we will denote
the set of all states on $\cA$. For any Hilbert space $\cH$ we denote
by $\cB(\cH)$ the set of all bounded linear operators on $\cH$.
Clearly, $\cB(\cH)$ is an example of $C^*$-algebra.

A linear map $\tau:\cA\to \cB$ between $C^*$-algebras is called {\it positive}
if $\tau(\cA^+)\subset \cB^+$. For $k\in \Nn$ we consider a~map
$\tau_k:M_k(\cA)\to M_k(\cB)$ where $M_k(\cA)$ and $M_k(\cB)$ are the algebras
of $k\times k$
matrices with coefficients from $\cA$ and $\cB$ respectively, and
$\tau_k([a_{ij}])=[\tau(a_{ij})]$.
We say that $\tau$ is {\it $k$-positive}
if the map $\tau_k$ is positive. The map $\tau$ is said to be
{\it completely positive}
when it is $k$-positive for every $k \in \Nn$.

The triple $(\cA, \tau, \omega)$ consisting of a unital $C^*$-algebra $\cA$, a linear
positive unital map $\tau$, and a state $\omega$ will be
called a {\it (quantum) dynamical system.}
We will need:

\begin{definition}
We say that the quantum dynamical system $(\cA, \tau, \omega)$ satisfies
detailed balance II if  there exists another linear positive unital map $\tau^{\beta}$
such that
\begin{equation}
\omega(A^*\tau(B)) = \omega(\tau^{\beta}(A^*)B)
\end{equation}
for all $A, B \in \cA$.
\end{definition}

There have been various versions of the detailed balance condition (cf discussion in \cite{M1}).
Here we would mention only the detailed balance I given in \cite{M2}
as that version is related to the existence
of a form of time-reversal for the underlying dynamics. The relations between both
conditions are described in \cite{M1}.

We will assume that the dynamical system $(\cA, \tau, \omega)$
satisfies detailed balance condition (DB)II. We recall that DB II
(the same will be true under DB I) implies: i) the state $\omega$
is $\tau$-invariant, ii) in the GNS representation $(\cH,
\pi_{\omega}, \Omega)$ of $(\cA, \omega)$, the definition
$T_{\omega} \pi_{\omega}(A) \Omega = \pi_{\omega}(\tau A) \Omega$
gives a contraction $T_{\omega}$ on the Hilbert space $\cH$, iii)
if additionally $\omega$ is a faithful state then $T_{\omega}$
commutes (strongly) with the associated modular operator.

Finally, to formulate the quantum Khintchin theorem we need the concept of a relatively dense subset.
We say that $\cN \subset \Nn$ is relatively dense provided that there exists an $L>0$
such that in any interval of natural numbers having length larger than $L$
one can find a number $n \in \cN$. Recently, the following quantum generalization
of Khintchin's theorem was proved (see \cite{Stroh}, \cite{Rocco})

\begin{theorem}
\label{czinczin}
Let $\cA$ be a $C^*$-algebra, $\varphi$ a state on $\cA$ and $\tau: \cA \to \cA$
a positive linear map such that $\varphi \circ \tau = \varphi$. Let us assume that
\begin{equation}
\label{schwarz}
\varphi(\tau(A)^* \tau(A)) \le \varphi(A^*A)
\end{equation}
for every $A \in \cA$. Then, for every $A \in \cA$ and $\epsilon >0$, there exists
a relatively dense subset $\cN$ of $\Nn$ such that
\begin{equation}
Re \varphi(A^* \tau^n(A)) \ge |\varphi(A)|^2 - \epsilon
\end{equation}
for all $n \in \cN$.
\end{theorem}

The aim of that note is to show that DB II combined with the
quantum Khintchin theorem (so with quantum recurrences) compel a
quantum dynamical system to pattern upon reversible evolution.
Clearly, it can be considered as a ``{\it quantum reminiscence}''
of the famous controversy between Boltzmann and Poincar\'e; $\tau$
in Theorem \ref{czinczin} represents a general stochastic map!
Here, we will argue that our result may be used in the study of
decoherence and stability of a large class of quantum systems.
Namely, defining decoherence to be an irreversible emergence of
classical properties in a quantum system (so disappearing of
macroscopic interferences) one can say that the essential
character of decoherence appears to be irreversibility (cf
\cite{Omnes}, \cite{Zurek}, \cite{Proc}, and \cite{Schlos}). In
other words, it seems that decoherence is not an intrinsic
property of Nature but rather a dynamical effect. In that context,
our next result says that DB together with quantum recurrences
spoil the stability of dynamics thereby producing an obstacle for
creation dynamical effects which could have lead to a decoherence
phenomenon.

\section{Stability}
As mentioned, the DB II implies that $T_{\omega}$ is a
contraction. Combining that result with the fact that
(\ref{schwarz}) was used in the proof of Theorem \ref{czinczin}
only to get a contraction in the GNS space one has
\begin{equation}
\label{konsek1}
||a \Omega|| ||T^n_{\omega} a \Omega|| \ge |(a \Omega, T^n_{\omega}a\Omega)|
\ge Re(a \Omega, T^n_{\omega} a \Omega) \ge |(\Omega, a \Omega)|^2 - \epsilon
\end{equation}
for any $a \in \pi_{\omega}(\cA) \equiv \cA_0$, and $n \in \cN$.
Suppose, $\omega(A) \ne 0$ and $\lim_{n \to \infty} ||T^n_{\omega} a \Omega|| = 0$
for $\pi_{\omega}(A) \equiv a \in \cA_0$.
Hence, $\forall_{\epsilon >0} \exists_N \forall_{n > N} \quad
||T^n_{\omega} a \Omega|| < \epsilon$.
This and (\ref{konsek1}) implies
\begin{equation}
\label{ala1}
\forall_{\epsilon >0} \exists_N \forall_{n:n> N \land n \in \cN} \quad \epsilon ||a\Omega||
> ||T^n_{\omega} a \Omega|| ||a \Omega|| \ge |(\Omega, a \Omega)|^2 - \epsilon
\end{equation}
which is a contradiction. Hence, for any $a (= \pi_{\omega}(A)$
such that $\omega(A) \ne 0$ the sequence $\{ ||T_{\omega}^n a
\Omega|| \}$ does not go to $0$. The limit exists as the sequence
being monotonic nonincreasing and bounded below is convergent.
Thus we get a form of stability for the discrete evolution $\{
T^n_{\omega} \}$.

Let us discuss the consequences of that result. Firstly, we recall
that for positive semigroups on $C^*$-algebras with unit, weak
stability and uniform stability coincide (see \cite{LNM}, Theorem
1.7 in Chapter D-IV). This means that we are not able to split our
original algebra into two subalgebras in such a way that there
would exist two $\tau$-invariant subalgebras, one of them such
that expectation values for observables from that set are
practically equal to zero after large time. The main obstacle to
such a splitting would be the existence of the time invariant
state that is guaranteed by the DB condition.

Therefore, one can expect a similar decomposition of observables
associated with the considered system to that given in  (cf.
\cite{BLO}, \cite{LO}) but now only in the representation space
$\cH$. We emphasize that such a type of decomposition was the main
ingredient of the discussion of decoherence in \cite{BLO},
\cite{LO}. Let us define the desired form of decomposition. The
search for a decomposition of the full algebra which contains
selected observables with a weak stability property can be
justified by the phenomenon called the environment-induced
decoherence (cf \cite{Schlos} and the references given there).
Assume $\cA$ is a $W^*$-algebra, so $\cA_0$ is a von Neumann
algebra on the Hilbert space $\cH$. Further, $\omega$ is a
faithful normal state given by a cyclic and separating vector
$\Omega \in \cH$. We wish to have:
\begin{equation}
\label{rozklad}
\cA_0 = \cA_1 \oplus \cA_2
\end{equation}
where both subsets $\cA_1$, $\cA_2$ are $\pi_{\omega}(\tau)$-invariant
and the following properties hold:
\begin{itemize}
\item $\cA_1$ is a von Neumann subalgebra of $\cA$ and the evolution $\pi_{\omega}(\tau)$
when restricted to $\cA_1$ is reversible.
\item $\cA_2$ is a linear space (closed in the norm topology) such that for any observable
$B=B^* \in \cA_2$ and any normal state $\phi$ of the system
with the support $s(\phi^{\prime})$
of its extension $\phi^{\prime}$ on $\cB(\cH)$
orthogonal to $|\Omega><\Omega|$, there is
\begin{equation}
\label{ala}
\lim_{t \to \infty}  \phi(\pi_{\omega}(\tau)(B)) = 0
\end{equation}
\end{itemize}

However we note that  (\ref{ala}) implies
\begin{equation}
\label{ala2}
\forall_{x \in \cA_0^{\prime}} \forall_{\epsilon >0}  \exists_{N = N(\epsilon, x)} \forall _{n>N}
\quad |(x^*x \Omega, T^n_{\omega} B \Omega)| < \epsilon
\end{equation}
where $B \in \cA_2$. Here, $\cA_0^{\prime}$ stands for the
commutant of $\cA_0$. We have also used that $\Omega$ is both a
cyclic and separating vector. Thus, an $f \in \cH$ can be
approximated by vectors of the form $\{ y \Omega,\quad y \in
\cA_0^{\prime} \}$. On the other hand, for all $B \in \cA_2$ such
that $\omega(B) \ne 0$, the conctractivity of $T_{\omega}$ and
(\ref{konsek1}) imply
\begin{equation}
 \forall_{\epsilon >0}  \exists_{N = N(\epsilon)} \forall _{n:n>N \land n \in \cN}
\quad const - \epsilon < ||T^n_{\omega}B \Omega || < const + \epsilon
\end{equation}
where $const$ is a positive number (depending on $B$).
This leads to

\begin{equation}
 \forall_{\epsilon >0}  \exists_{N = N(\epsilon)} \forall _{n:n>N \land n\in \cN} \exists_{f \in \cH}
\quad const - 2\epsilon < |(f,T^n_{\omega}B \Omega )| < const + 2\epsilon
\end{equation}
which would contradict (\ref{ala2}). We have used that
\begin{equation}
\forall_{\epsilon >0}  \exists_{f \in \cH} \quad | ||T^n_{\omega} B \Omega||
- (f, T^n_{\omega} B \Omega )| < \epsilon.
\end{equation}
Consequently, only a very specific decomposition of the von Neumann algebra associated to
the set of observables of the system will be possible.

\section{Discussion}
Inequality (\ref{schwarz}) is nothing but a composition of the
Kadison-Schwarz inequality with a state. To get it $2$-positivity
would be enough. However, to get the decomposition of the form
(\ref{rozklad}) one needs some extra conditions (cf \cite{LO}). It
is an easy observation that DB provides these conditions. In other
words, a general dynamical semigroup consisting of completely
positive maps does not fulfil the necessary requirements unless it
possesses additional properties, eg. the DB condition.

Further, as we do not expect that the collective variables form a
$C^*$-subalgebra we should distinguish between weak and uniform
stability. Clearly, this makes sense if the considered system has
an infinite number of degrees of freedom. At this point it is
worth mentioning that it was Heisenberg who pointed out, on
various occasions, the role of environment (so infinite systems)
in the problem of suppression of macroscopic interferences.
Therefore, our approach relying on the idea of infinite systems is
well justified.

Our approach sheds some new light on
relations between recurrences and stability of dynamical semigroups $V_t$
on a Hilbert space ($V_t$ is a one parameter strongly continuous semigroup of linear
contractions on $\cH$). Namely, strong stability of such semigroups is defined as
\begin{definition}
The semigroup $V_t$ on $\cH$ is strongly stable if as
$t\to\infty$  $\Vert V_tf\Vert \to0$ for all $f\in{\cH}.$
\end{definition}
\smallskip
Let us recall (cf. \cite{[Go]})
\smallskip
\begin{theorem}
Let the semigroup $V_t$ be a contraction semigroup
on ${\cH}$. ${\cH}$ has a maximal closed subspace
${\cH}_1$ on which $V_t$ is (i.e. restricts to) a unitary
semigroup. The restriction of $V_t$ on ${\cH}^{\bot}_1$
is a completely non unitary semigroup. Moreover, both
$V_t$ and $V^*_t$ are strongly stable on ${\cH}^{\bot}_1$
if and only if $P=Q$ is a projection, where
\begin{equation}
\label{gol}
Pf=\lim_{t\to +\infty}V^*_tV_tf \quad \& \quad Qf = \lim_{t\to +\infty}
V_tV^*_tf
\end{equation}
for $f\in{\cH}$. The range of $P=Q$ is then ${\cH}_1$.
\end{theorem}

\smallskip
\begin{remark}
\item{(i)} The limits in (\ref{gol}) exist (cf. \cite{[FSz-N]}).
\item{(ii)} Conditions leading to semigroups being strongly stable
on ${\cH}^{\bot}_1$ was also studied in (\cite{[M3]} and \cite{[M4]}).
\end{remark}
\smallskip
Stability of semigroups is also strongly related to peculiar properties of infinitesimal
generators of semigroups. Namely,
\begin{theorem}(see \cite{[AB]})
 Let $V_t$ be a bounded $C_0$-semigroup with generator $A$.
Assume that ${\sigma}_r(A){\cap}i{\bf R}=\emptyset$,
where ${\sigma}_r(A)$ denotes the residual part of the
spectrum of $A$. If ${\sigma}(A){\cap}i{\bf R}$ is
countable, then $V_t$ is a strongly stable $C_0$-semigroup.
\end{theorem}
\smallskip

Firstly, we recall that a dynamical semigroup $\tau_t$ (one
parameter semigroup of linear unital positive maps on $\cA$)
satisfying DB II or DBI gives rise to a dynamical semigroup of
type $V_t$ on the GNS Hilbert space $\cH$ such that $\sigma_r(A) =
\emptyset$ if DBI holds. Secondly, the results of Section 2 have
shown that the quantum recurrences spoil stability. Consequently,
the recurrence phenomenon leads to a peculiar form of the spectrum
of the infinitesimal generator of $V_t$ (uncountable intersection
of the spectrum with the imaginary axis) and gives special
asymptotic behaviour of the semigroup ($P \ne Q$, cf (\ref{gol})).
Furthermore, for dynamics consisting of positive maps satisfying
DB II there is only room for ``scattered'' collective observables,
i.e. they do not form such a ``rich'' structure as that which was
used to define $\cA_2$. Consequently, one may conjecture that the
interaction between the system and the environment singles out a
subset of states with well defined stability for some but not all
of the operators. 

It may be worth noting that the obtained results could be viewed 
from another perspective.
If one takes stability of certain observables in selected states as a given, then 
the contradiction obtained in (2.7) can also be interpreted as
saying that recurrence of quantum phenomena in the sense of Stroh, Zsido, et al,\cite{Stroh},
can 
only occur in states which suppress stability in the sense of annihilating
$\mathcal{A}_2$. Hence in the presence of the DB condition, stability limits recurrence.

\end{document}